\documentclass[structabstract,printer]{aa} 
\usepackage{graphicx}
\usepackage{natbib}
\usepackage{amsmath}
\usepackage{txfonts}
\usepackage{color}

\newcommand{\be}{\begin{eqnarray}}
\newcommand{\ee}{\end{eqnarray}}

\newcommand {\nbodypp}{\textsc{\mbox{nbody6\raise.4ex\hbox{\tiny++}}}}
\newcommand {\Msun} {\mbox{M$_{\odot}$}}

\newcommand {\SFE} {$\epsilon_{{\mbox {\scriptsize SFE}}}$}

\begin{document}

\title{Reaction of  Massive Clusters to Gas Expulsion -\\
The cluster density dependence}
\author{S. Pfalzner\inst{1} \& T. Kaczmarek\inst{1}}
\institute{
\inst{1}Max-Planck-Institut f\"ur Radioastronomie, Auf dem H\"ugel 69, 53121 Bonn, Germany\\
\email{spfalzner@mpifr.de}}
\date{ }

\abstract
   {The expulsion of the unconverted gas at the end of the star formation process potentially leads to  the expansion of the just formed stellar cluster and membership loss.  The degree of expansion and mass loss depends largely on the star formation efficiency and scales with the mass and size of the stellar group as long as stellar interactions can be neglected. }
   {We investigate under which circumstances stellar interactions between cluster members become so important that the fraction of bound stars after gas expulsion is significantly altered. }
   {The Nbody6 code is used to simulate the cluster dynamics after gas expulsion for different SFEs. Concentrating on the most massive clusters observed in the Milky Way, we test to what extend the results depend on the model, i.e. stellar mass distribution, stellar density profile etc., and the cluster parameters, such as cluster density and size.}
   {We find that stellar interactions are responsible for up to 20\% mass loss in the most compact massive clusters in the Milky Way, making ejections the prime mass loss process in such systems. Even in the loosely bound OB associations stellar interactions are responsible for at least $\sim$ 5\% mass loss. The main reason why the importance of encounters for massive clusters has been largely overlooked is the often used approach of a single-mass representation instead of a realistic distribution for the stellar masses.  The density-dependence of the encounter-induced mass loss is shallower than expected because of the increasing importance of few-body interactions in dense clusters compared to sparse clusters where 2-body encounters dominate.}
    {}

\maketitle

\section{Introduction}

Star clusters\footnote{A brief comment on terminology: Some authors use the term "cluster" to refer only to stellar groups that remain bound after gas expulsion, while other authors use it for any significant stellar over-density regardless of its dynamical state. Here we use the word "cluster" in the latter sense.}
are initially still embedded in the gas and dust they are forming from. The conversion from gas and dust to stars being incomplete, the left over gas/dust component is driven outward by stellar feedback. For embedded clusters containing more than a few hundred stars this can come in the form of photo-ionizing radiation, the winds of high-mass stars, and the onset of the first supernova explosions (Goodwin 1997).   

The subsequent response of the star cluster to the impact of gas expulsion has been the subject of a large number of theoretical investigations (Tutukov 1978, Hills 1980, Lada, Margulis  \& Dearborn 1984, Verscheuren \& David 1989, Goodwin 1997, Kroupa, Petr  \& McCaughrean 1999, Adams 2000, Geyer  \& Burkert 2001, Kroupa, Aarseth  \& Hurley 2001, Boily  \& Kroupa 2003a,b, Fellhauer  \& Kroupa 2005, Bastian \& Goodwin 2006, Banerjee \& Kroupa 2013).  

Already Hills (1980) found that the star formation efficiency \SFE\ of the cluster, i.e. the fraction of gas that is converted into stars,
\be \epsilon_{{\mbox {\scriptsize SFE}}} = \displaystyle{\frac{M_{st}}{M_{st}+M_{gas}}},\ee
where $M_{st}$ is the stellar mass and $M_{gas}$ the gas mass, is a key property in determining the fate of the cluster after gas expulsion.   Later numerical modelling (Lada et al. 1984, Geyer  \& Burkert 2001, Boily  \& Kroupa 2003a,b) showed that the actual limit for bound star cluster formation in case of instantaneous gas loss is in the range of  0.25 $<$ \SFE $<$ 0.4.

In the solar neighbourhood the majority of embedded clusters seem not to develop into longer-lived open clusters  (Lada \& Lada 2003, Porras et al. 2003), but disperse their stars early on ($<$10 Myr) feeding the field star population. This is largely attributed to the star formation efficiency (SFE) being too low that the cluster could withstand rapid gas loss (Goodwin 2009; Baumgardt \& Kroupa 2007). This is in accordance with observations which find typical SFEs in the solar neighbourhood, to lie in the range  0.1 $<$ \SFE $<$ 0.35 (Lada 1999, Lada  \& Lada 2003). For an alternative explanation for the high cluster infant mortality see Smith et al. (2011). 

There are strong indications that in starburst clusters, preferentially located close to the Galactic centre and in the spiral arms, the SFE could be higher ($\geq$50\%). In these systems infant mortality might be lower and gas expulsion less important (Bastian 2012). 
Given the large uncertainty in observed SFEs and the strong indications that star formation efficiencies depend on the local gas densities in the molecular cloud (Gutermuth et al. 2011),  we will cover the entire spectrum from \SFE=0.1 to \SFE=1.0 in the here presented simulations.

Extensive parameter studies of the effect of gas expulsion on {\em massive} clusters ($N>$10$^4$) are rare,  as the computational effort is proportional to $N^2$, where $N$ is the number of simulated stars.  One noteworthy exception is the parameter study by Baumgardt \& Kroupa (2007). To minimize the computational cost some approximations, namely the use of a single mass representation for the cluster stars and a profile that is not extremely centrally condensed, are utilized  that lead to fewer stellar dynamical interactions. 

Apart from the SFE other factors determine as well the
bound fraction after gas expulsion.  So can longer gas expulsion timescale, higher cluster concentrations, sub-structured initial conditions, subvirial velocities, and radially decreasing SFEs
(Lada et al. 1984, Goodwin 1997a, Adams 2000, Fellhauer  \& Kroupa 2005, McMillan et al. 2007, Offner et al. 2009, Allison et al. 2011, Smith et al. 2013)  lead to clusters of \SFE$<$ 0.3  still retaining bound cluster remnants. 
In the present study we take none of these effects into account, since the aim is to isolate
a process that has so far been largely overlooked, namely, the importance of stellar dynamical interactions.

We will show that using a single-mass representation was one of the main reasons why the importance of dynamical encounters has so far been underestimated. Although there were some first hints that encounters might be important in the development of young clusters (Stahler \& Converse 2010, Allison et al. 2010), only recently Moeckel et al. (2012) found that using single masses hides the effect of encounters when modelling cluster development after gas expulsion in small $N$ clusters. They attribute the importance of encounters to the smaller ratio of the relaxation to crossing time in small $N$ clusters in comparison to massive clusters.

By contrast to the study by Moeckel et al. (2012) we concentrate on massive clusters typically containing $>$ 10$^4$ stars and perform the first quantitative evaluation of the importance of encounters as a function of the cluster density. 

The aim of this paper is to demonstrate that the stellar interaction dynamics play an important role in the early expansion of typical massive star clusters ($>$ 5 $\times $10$^3$ \Msun) of the Milky Way. After describing the numerical method in section 2, it is demonstrated in section 3 why this effect has been largely overlooked in the past. In section 4 the results of the extensive parameter study are presented and summarised by a formula for the bound fraction after gas expulsion including cluster density effects.

\section{Method}

We use the code Nbody6 (Aarseth 2003) to perform an extensive parameter study of the dynamics of massive star clusters ($N>$ 10$^4$) after gas expulsion. For such massive clusters the gas expulsion time scale is so short that it can be modelled as instantaneous (Fellhauer \& Kroupa 2005). Thus the method described by Goodwin \& Bastian (2006) of not explicitly modelling the gas expulsion process itself, but using the equivalent supervirial representation at the start of the simulation can be applied here. To test the validity of this approach, we modelled as well the full process of gas expulsion on short time scales  for a subset of the parameter space using a temporale-dependent background potential and found no difference in the results.

\begin{table}
\begin{center}
\begin{tabular}{r *{6}{c}}
ID & profile & $M_s$  & No. stars & $r_{hm}^i$ & $N_{sim}$ & error(\%)\\[0.5ex]
\hline
P0     & Plummer & single  & 30 000      & 1.3  &  15  & 0.5 \\
P1     & Plummer & IMF      & 30 000      & 1.3  &   15 & 2-3\\
LK0   & King        & single  & 30 000      & 1.3  &   15 & 2-3\\
LK1   & King       & IMF      & 30 000      & 1.3  &   15  & 2-3\\
LK2   & King       & IMF      & 30 000      & 4.6  &   15  & 2-3\\
LK3   & King       & IMF      & 45 000      & 1.3  &   15  & 2-3\\
LK4   & King       & IMF      & 15 000      & 1.3  &   15  & 2-3\\
CK1    & King       & IMF       & 30 000      & 0.1 &     7  & 3-4 \\
CK2    & King       & IMF       & 30 000       & 0.2 &    5  & 3-4\\
CK3    & King       & IMF       & 30 000      & 0.3 &     5  &3-4\\ 
CK4    & King       & IMF       & 30 000       & 0.5 &    5 &3-4\\

\end{tabular}
\caption{Properties of the modelled clusters, where ID stands for the identifier, the second column shows the shape of the used cluster profile, the in the third column $M_s$ for the distribution of stellar masses where IMF implies the Kroupa 2001 version, the forth column the number of stars, the fifth column shows the initial half-mass radius in pc, sixth and seventh columns show the number of simulations performed for the given set-up and the resulting standard deviation. 
\label{table:sim}}

\end{center}
\end{table}

The cluster parameters are chosen in such a way that they span the entire range of properties observed for massive clusters at ages less than 4 Myr in the Milky Way (Pfalzner 2009, Portegies Zwart et al. 2010). In Table 1 the parameter space of our study is detailed.
For each of these parameter combinations we model clusters with \SFE= 0.1, 0.2, 0.3, 0.4, ..., 1.0.

Using GPU-based computers speeds up the simulations to such an extend that systems with a large stellar membership can be modelled with high statistical significance by averaging over a larger number of cluster representations. The last two columns of Table 1 give the number of realisations and the corresponding standard deviation. 

In models P0 and LK0 a single-mass representation for all stars was used. In all other models the stellar masses are distributed according to the IMF given by Kroupa (2001).
The star are spatially distributed according to a Plummer profile (Plummer 1911) in models P0 and P1, and according to a King profile (W0 = 9) (King 1966)  in all other models. 

Both models were widely used in the past to model young clusters, although they are really only representative of the stellar density distribution in old globular clusters. Comparison with observations show that young clusters are better represented by King models with $W_0 >$7 (for example, Hillenbrand \& Hartmann 1998).  In the following the term ``King model'' is used as equivalent to King distributions with high $W_0$ values.

In our simulations we apply  two simplifications to reduce the computational expense - we neglect stellar evolution and start without primordial binaries. These assumptions will be critically discussed in section 6. However, an equally extensive quantitative parameter study including these two effects, especially with realistic primordial binary fraction of $\geq$ 50\%, seems with current computational facilities difficult to achieve. Even for the least dense systems including primordial binaries increases the computation time by a factor $\sim$ 10, for the dense systems (like, for example, model CK1) simulations can last easily a factor 100 or even 1000 longer.

Tidal disruption was not included in this study. Apart from clusters located close to the Galactic Centre this effect only plays a significant role at later times than the here investigated first 20 Myr of cluster development.

\section{Model dependence}

In the following we want to investigate how the bound fraction after gas expulsion depends on the cluster model used.

\subsection{Stellar mass representation}

\begin{figure}[t]
\includegraphics[width=0.5\textwidth]{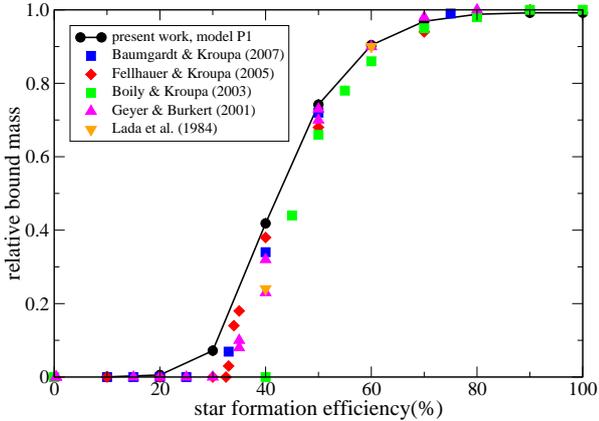}
\caption{Overview of results from the literature of the relative bound mass as a function of SFE. The solid line represents our results for a cluster with a Plummer distribution of single-mass stars (model P0).}
\label{fig:other_results}
\end{figure}

Figure \ref{fig:other_results} summarises the results of previous work by Lada et al. (1984), Geyer \& Burkert (2001), Boily \& Kroupa (2003), Fellhauer \& Kroupa (2005) and Baumgardt \& Kroupa (2007). It shows the bound mass fraction as a function of the star formation efficiency. Each of these models exclude encounters between cluster members either explicitly  by their simulation method (gravitational softening) or implicitly by using a single mass representation of the stellar population. The general trend in all these simulations is similar: For \mbox{$\epsilon_{SFE} \leq $20\%} basically the entire cluster dissolves, for   $\epsilon_{SFE} \sim $30\% a remnant cluster remains, however, it contains only $\approx$ 5\%-10\% of the initial mass. For higher SFEs the mass of the remnant cluster increases steeply with the SFE.

Despite the general trend in these results being very similar, some discrepancies are found for SFEs between 20\% and 40\%. The most likely explanation for these differences is the statistical nature of the results. In the past, preferentially clusters with a membership of a few thousand stars were modelled and
simulations were, if at all, repeated only a few times. Fig. \ref{fig:other_results} shows in addition our results of model P1, where we use 30 000 particles and perform 15 simulations per SFE. Due to higher statistical significance the error is only $\sim$ 0.5\%. If we use our error analysis to estimate the error bars in previous simulations, we obtain errors of up to 10\%. Thus the differences in Fig. \ref{fig:other_results} can indeed be largely attributed to statistical fluctuations.

\begin{figure}[t]
\includegraphics[width=0.45\textwidth]{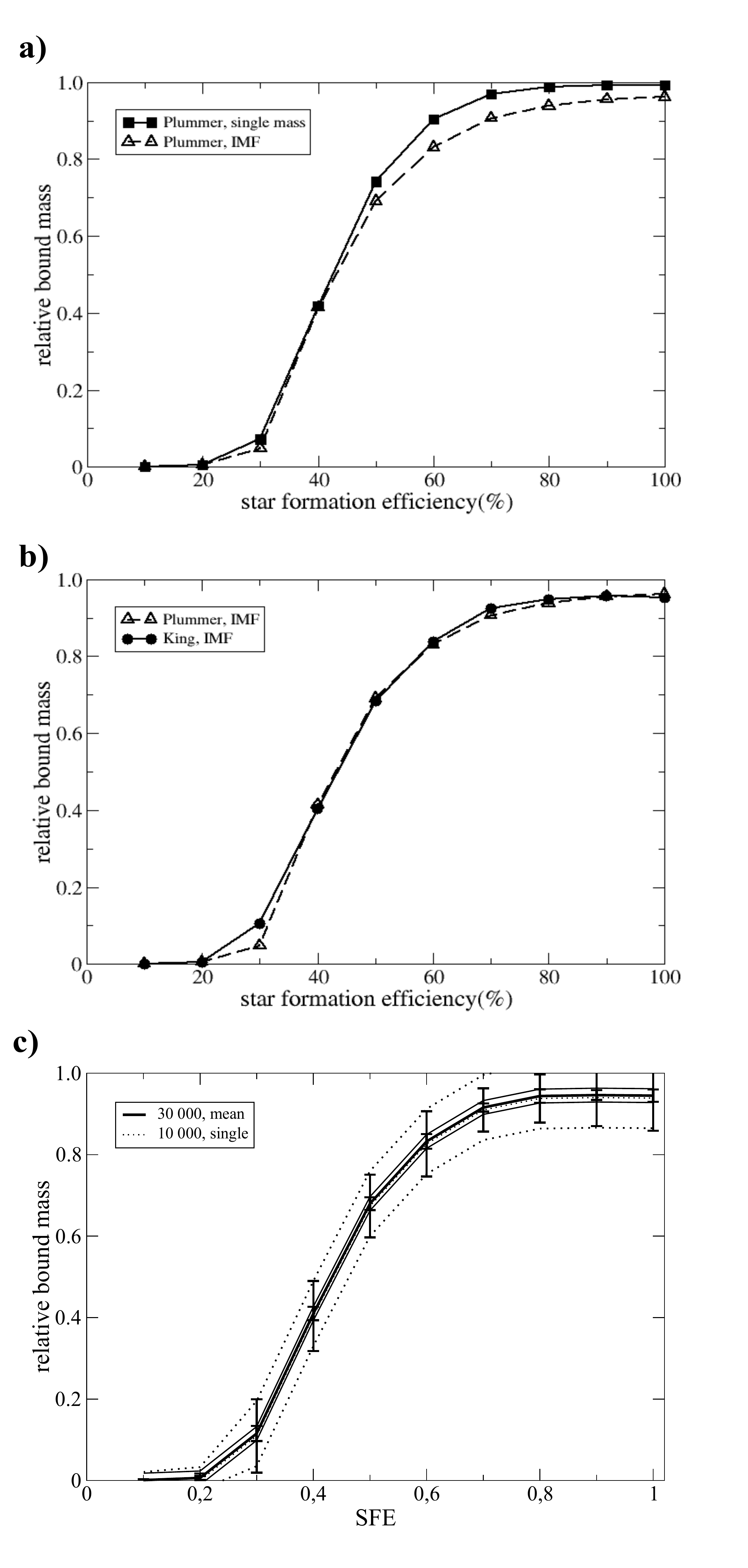}
\caption{Relative bound mass at 20 Myr as a function of SFE. a) shows a comparison of cluster models with a Plummer distribution of single stars (model P0 - solid line, squares) and stars chosen according to an IMF (model P1 - dashed line, triangles). b) shows the comparison between cluster models with a Plummer (model P1 - dashed line, triangles) and a King distribution (model LK1 - solid line, circles. c) shows the results for model LK1 with 30 000 stars (thick solid line), but this time with the error bars for results averaged over 15 simulation runs (thin solid line) and the error a single simulation with 10 000 stars would yield (dotted line).
}
\label{fig:comparison}
\end{figure}

We repeated the same type of investigation, but used a stellar population according to the IMF of Kroupa (2001) (model P1). Although the general trend is similar,  Fig. \ref{fig:comparison}a) shows an important difference: for  $\epsilon_{SFE} > $70\% the simulations with single masses (model P0) show basically no mass loss whereas the simulation with masses according to the IMF (model P1) show $>$5\% mass loss. This mass loss is due to encounters between cluster members leading to ejections. Obviously even in the here considered case, which corresponds to the high-mass end of the relatively wide spread OB associations/loose clusters (see paper 2) encounters lead to a non-negligible cluster mass loss. We conclude that the often used simplification of modelling the effect of gas expulsion by using single stellar mass models hides this important effect of massive clusters and association dynamics.

\subsection{Spatial distribution}

Next we investigate to what extend the chosen cluster profile
influences the result. In Fig. \ref{fig:comparison}b)  the dashed line shows the results for Plummer-distributed (model P1), whereas the drawn line denotes King-type clusters (model LK1). All other model parameters are the same. 
The general trend is again very similar,  even the loss caused by stellar interactions is nearly as
high in the case of a Plummer-type cluster as for a King profile. However, for \SFE\ = 0.3 the bound mass for Plummer-type clusters is only about half that of King-type clusters. The reason is that Plummer profiles show a much lower concentration of stars in the cluster centre. When a cluster expands as a response to gas expulsion, mass loss occurs predominantly from outside inwards. Consequently, as in King-shaped clusters more mass is concentrated in the central regions, the remnant cluster mass is higher. This effect is most pronounced for SFEs where the remnant cluster basically consists of the former inner core of the cluster, which is the case for $\sim$ 30\% SFE.

\subsection{Statistical significance}

One reason why the significance of encounters and the influence of the cluster profile in the 30\% SFE range was underestimated in the past, is that often only a small sample of simulations was considered. Fig.\ref{fig:comparison}c shows the error considering a single 10 000 star simulation, a single \mbox{30 000 } star simulation and the average over 15 simulations of clusters containing \mbox{30 000} stars. 

For a single simulation with 10 000 stars it would be equally likely to conclude that for a 100\% SFE case one would obtain no mass loss or a 15\% mass loss. Equally such large errors allow for the case of \SFE\ = 0.3, a bound fraction of 2\%-20\%. So the above considered effects can easily hide within the statistical bandwidth. Performing 15 realisation of a 30 000 star cluster model reduces the error to 2-3\%. Only with such a small error is it possible to detect 
the 5-10\% mass loss due to encounters in this type of cluster and that Plummer-profiles give smaller bound fractions than King-shaped clusters for intermediate SFEs.

For quantitative results with errors $<$3\% one needs to follow at least 500 000 stars over the course of a set of simulations. We achieved such an accuracy for all
investigated parameters, apart from simulations of extreme dense clusters (models CK1 - CK4). Here the $\sim$ 1000 time higher density leads to much longer computation times due to the high number of close encounter. In addition, the number of simulations that stop before 20 Myr due to numerical problems increases as well. 

One might wonder why the error  for model LK1 is about 2-3\%  (Fig. \ref{fig:comparison}c) whereas the error for model P1 is $<$1\%. The reason is  the presence of encounters in model LK1. The occurrence of encounters is strongly determined by the statistical fluctuations in the special distribution of the cluster members which depends on the details of the upper end of the mass spectrum and the number and nature of binaries present.

We find that mostly low-mass stars are ejected. This holds as well in relative terms:  ejections are responsible for $\sim$6\% mass loss from all stars, but only 2\% for B-type stars in model LK1.

\section{The influence of cluster density}

\begin{figure}[t]
\includegraphics[width=0.5\textwidth]{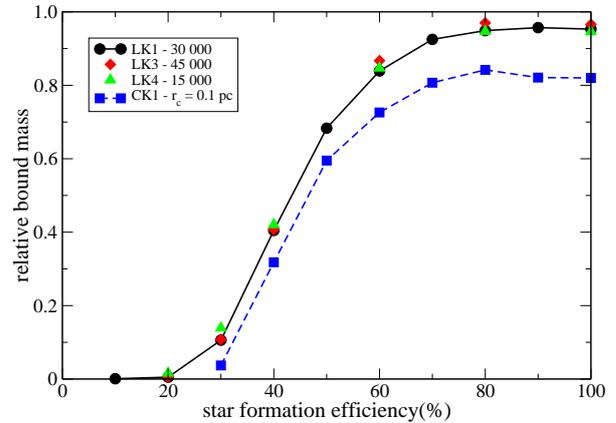}
\caption{The bound mass fraction after 20 Myr as a function of the SFE. The different symbols indicate the different cluster models as detailed in table 1. }
\label{fig:sfe-dens}
\end{figure}

Naturally the role played by encounters is a function of cluster density.
Fig. \ref{fig:sfe-dens} shows the bound mass fraction 20 Myr after gas expulsion as a function of the SFE, but this time for 15 000, \mbox{30 000}, and 45 000 stars (models LK4, LK1, and LK3). As each model has the same half-mass radius of $r_{hm}$ = 1.3 pc, the stellar density in model LK3 is 3 times higher than in model LK4. Nevertheless the results are basically the same considering the error of 2 - 3\%.

Only for 30\% SFE the result for less dense model LK4 shows a  slightly higher bound mass fraction than the denser models LK1 and LK3. The reason is not, as one might expect, the lower density in model LK4, but the longer crossing and relaxation time in the lower density cluster.  System LK4 has not yet reached its new equilibrium state at 20 Myr yet (see Parmentier \& Baumgardt 2012). Looking at later times, confirms that although it takes longer for model LK4 to reach the new equilibrium state, the bound mass fraction is then basically the same as in models LK1 and LK3. So a factor of three in density does not change the results in a perceivable way. 

\begin{figure}[t]
\includegraphics[width=0.47\textwidth]{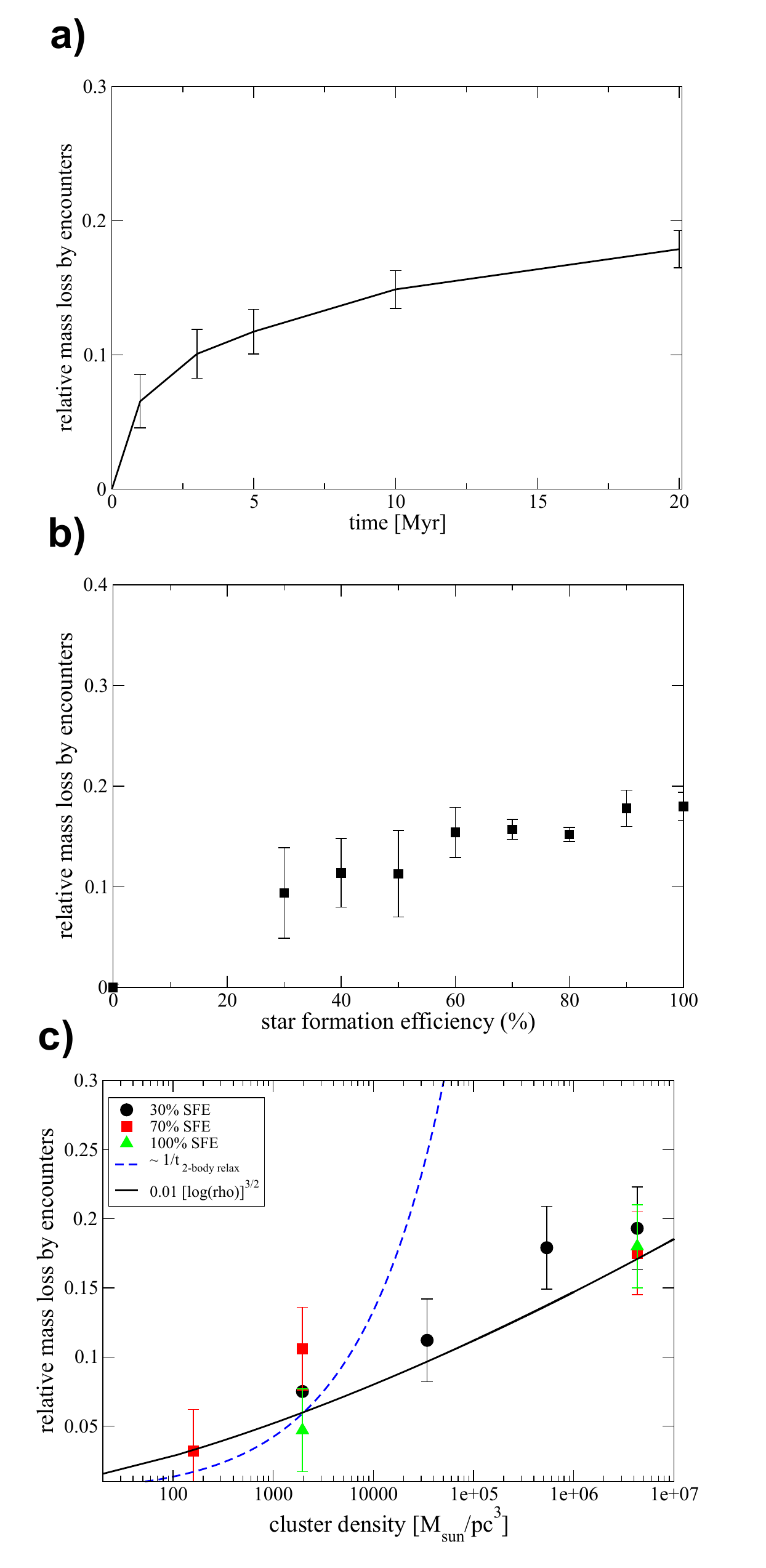}
\caption{Relative mass loss induced by stellar encounters a) as a function of time for model CK1 for the case of \SFE\ = 1.0, b) as function of the SFE for model CK1 , and c) as a function of the initial cluster density. In b) no data points for SFEs of 10\% and 20\% are plotted, as here the bound masses are so low and the errors are large, that no meaningful values for the differences can be obtained.
In c)  the different symbols give the values obtained from the simulation results for 
\SFE\ =0.3 (circle), \SFE\ =0.7 (square), and \SFE\ =1.0 (triangle). The curve shows the approximation 
of $\left(\log (\rho)\right)^{3/2}$.}
\label{fig:enc-time}
\end{figure}

Only if one goes to much denser clusters, as, for example, in model CK1, the dependence on the cluster density becomes apparent. The much smaller half-mass radius of $r_{hm}$ = 0.1 pc means a $\sim$ 2000 times higher stellar density. Naturally encounters become much more important in such an extremely dense environment.

First we compare the hypothetical case of 100\% SFE, which allows to study the impact of encounters without the gas expulsion process. Fig. \ref{fig:enc-time}a)  shows that for 100\% SFE the mass loss -  in this case completely due to encounters - is $\sim$ 18\% after 20 Myr of cluster development. This loss is not instantaneous but proceeds over the entire shown time span of 20 Myr (see Fig. \ref{fig:enc-time}a) and continues to some extend even afterwards. The steady but slow mass loss due to stellar interactions means that the cluster has enough time to respond by expanding until it reaches a new equilibrium state.

The mass loss due to stellar interactions is most pronounced during the first 3 Myr and decreases over time (see Fig.  \ref{fig:enc-time}a). The reason is that even for \SFE\  =1.0 the cluster CK 1 expands by approximately a factor 9 over 20 Myr leading to a decrease in cluster density by a factor of $\sim$ 700. Thus, stellar interactions become rarer with time.

For lower SFEs gas expulsion leads to rapid cluster expansion and a fast decrease in cluster density. Therefore one would expect a reduced effect of encounters and one could indeed interpret Fig. \ref{fig:enc-time}b) in this way. It shows the mass loss due to stellar interactions as the function of the SFE. However, for small SFEs the error is large because of the small number of stars in the remnant cluster.  Thus the data could as well be interpreted as being constant.  For simplicity we use this interpretation for the remainder of this investigation. 

Using different initial cluster radii (models LK2, CK2, CK3, and CK4 ) and performing the equivalent set of simulations, we determine the dependence of the relative mass loss due to stellar encounters on the initial cluster density. This is easiest for \SFE\ =1.0 where the mass loss is solely due to stellar encounters.  However, both mass loss processes can be separated as well for lower SFEs by subtraction from a very low density case. Fig. \ref{fig:enc-time}c shows that the relative mass loss due to stellar interactions is a weak function of the initial stellar density. This means, one needs at least a factor of 100 difference in density to detect this density dependence given in addition the large statistical scatter of the results. A fit formula for the bound mass fraction after 20 Myr as a function of the SFE including its density dependence is given in the appendix.

\section{Nature of the encounter-induced mass loss}

 Such a weak density dependence is somewhat surprising as according to classical theory one would attribute the mass loss due to encounters to 2-body relaxation processes. In this picture the rate of loss is given as (Binney \& Tremaine 2008) 
\be
\frac{dN}{dt} \propto - \frac{N}{t_{evap}}, 
\ee 
where the evaporation $t_{evap}$ is a multiple of the relaxation time $t_{relax}$, which in turn is proportional to the crossing time $t_{cross}$. In Fig. \ref{fig:enc-time}c all investigated clusters are of the same mass, therefore it would follow  for the relative mass loss through 2-body relaxation $\left( \Delta M/M \right)_{2-body}$
\be
\left( \frac{\Delta M}{M}  \right)_{2-body} \propto \frac{1}{t_{evap}} \propto \frac{1}{t_{relax}} \propto \frac{ln M}{\sqrt{(Mr_{hm}^3)}}\propto \rho ^{1/2}
\ee 
In  Fig. \ref{fig:enc-time}c this $\rho ^{1/2} $-dependence is plotted as dashed line. It can be seen that classical theory  based on 2-body relaxation processes predicts  a much stronger dependence  on the cluster density for very dense clusters than found in the simulations. One could now anticipate that perhaps the fraction of formed binaries or the degree of mass segregation is responsible for this deviation from the 2-body relaxation approach. 

Although all these processes might influence the result to some degree, we find that a change in the nature of the encounter dynamics at very high densities is the main reason for the observed discrepancy. Such a change was already noticed in the context of the influence of encounters on protoplanetary discs (Olczak et al. 2010, Duke \& Krumholz 2012, Olczak et al. 2012, Pfalzner 2013).
There it was found that at mean cluster densities of $\approx$10$^3$ pc$^{-3}$ parabolic encounters with low-mass stars ($<0.5$ \Msun) and the few most massive stars of the system dominate.
We find a similar transition in our simulations for the encounter induced mass loss in the gas expulsion phase.

\begin{figure}[t]
\includegraphics[width=0.5\textwidth]{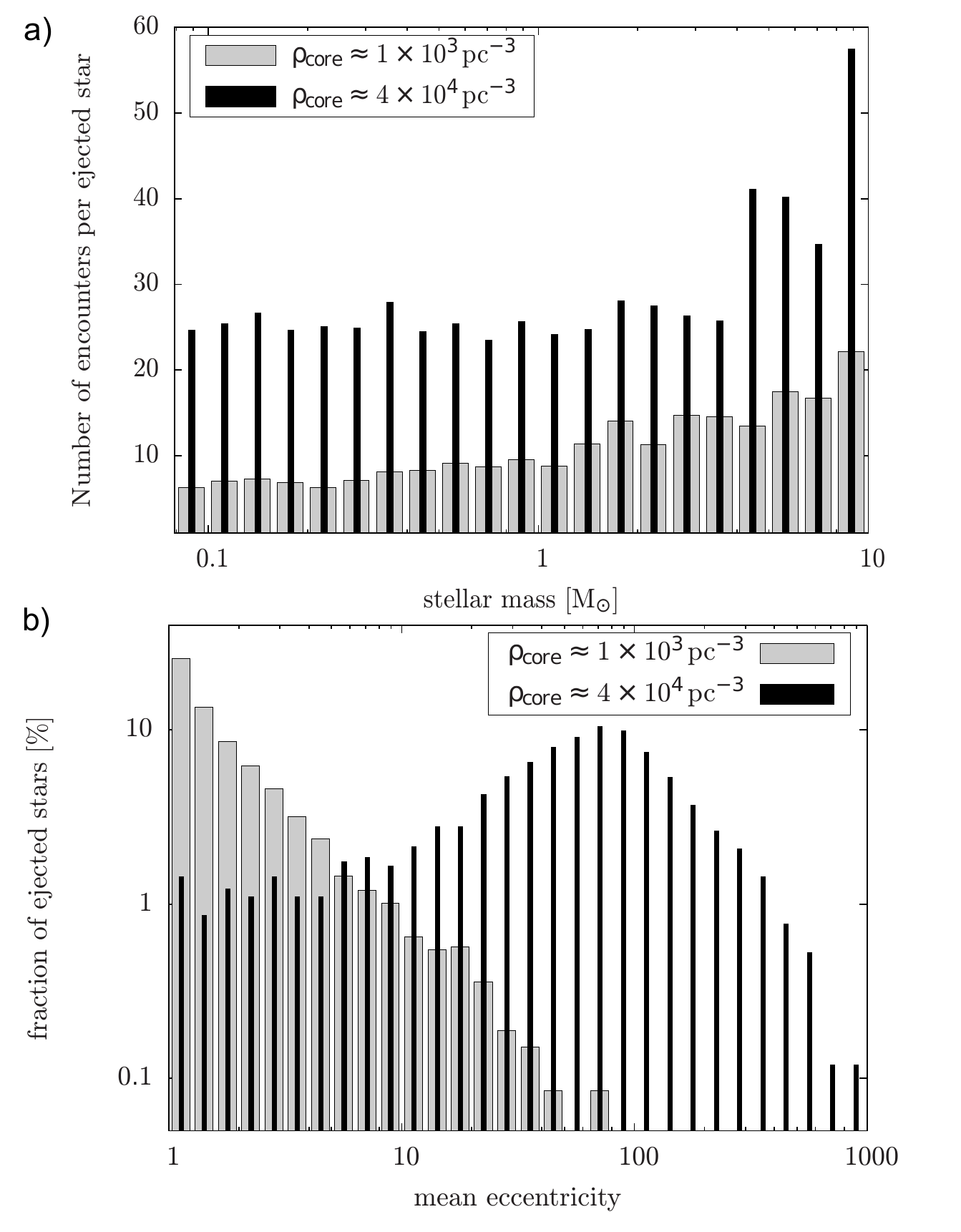}
\caption{a) Average number of encounters each star has before it becomes ejected as a function of stellar mass. The grey indicates clusters with $\rho$= 1 $\times$ 10$^3$ pc$^{-3}$ and black clusters with  $\rho$=4$\times$ 10$^4$ pc$^{-3}$.  b) shows for all stars that become ejected their mean eccentricity in the encounters that eventually lead them to become unbound.}
\label{fig:num_enc}
\end{figure}

In order to pin down the reason for this effect without the difficulty of the density changig during the simulation, we re-analysed our results described in Olzak et al. (2010), where the cluster density is relatively constant. We found that although the relative number of ejected stars increases in dense clusters, that at the same time the number of encounters per star before ejection increases compared to lower mass clusters (see Fig. \ref{fig:num_enc}a). The reason is that in contrast to low-density clusters, 
where most encounters are on parabolic orbits, in dense clusters highly hyperbolic encounters with the eccentricities ($e \gg 10$) (see Fig. \ref{fig:num_enc}b)) dominate.  

The reason for the dominance of hyperbolic encounters in very dense clusters is that there the density becomes so high that any encounter between two stars  is to some degree perturbed by the remainder of the cluster.  This means that the approximation of 2-body relaxation becomes invalid and is replace by few-body encounters. These few-body interactions are not necessarily encounters between a single star and a binary or higher order systems, but includes groups of unbound stars as well.

The resulting hyperbolic orbits are much less efficient in producing escapers than the equivalent parabolic 2-body encounters. Therefore it follows that
\be t_{relax}^{few body} \gg t_{relax}^{2-body},\ee
where $t_{relax}^{few body}$ is the few-body relaxation time. So the transition from predominatly 2-body encounters to few-body encounters in very dense clusters is the reason for the weaker than expected dependence on the cluster density of the encounter-induced mass loss. 

It is difficult to pin-point an exact cluster density where this transition to a few-body 
encounter dominated system happens from above simulations as due to the gas expulsion process the cluster density changes rapidly with cluster age. However, from our simulations described in Olczak et al. (2010) we can give a first estimate that the transition from 2-body to few body dominated encounters happens for mean cluster densities of $\approx$10$^4$ pc$^{-3}$. The encounters that lead to ejections mainly take place in the central cluster areas, where the density is 10 to 100 times higher than this average.

Taking above estimate as a guide, in the here investigated sample there are clusters that are throughout the gas expulsion phase 2-body dominated (model LK2), clusters that are initially few-body dominated but become 2-body dominated in the expulsion phase (model LK0) and models that largely stay in the few-body dominated regime throughout (model CK1). For the latter it explains as well the weak dependence of the encounter-induced mass loss on the SFE, despite the strong dependence of the expansion history on the SFE.

\section{Discussion}

In the present parameter study we made several commonly used approximations: gas expulsion was assumed to be instantaneous, neither sub-structuring, primordial mass segregation, nor stellar evolution were considered and the simulations contained no primordial binaries. The first two assumptions, instantaneous gas expulsion and no sub-clustering - are probably justified for the high-mass cluster considered here.  The large number of massive stars should lead to rapid gas expulsion and sub-structuring caused by the star formation process is probably quickly removed. 

Mass segregation can only change the encounter-induced mass loss in case gravitational focussing plays a role. So results of the dense end of our sample should not be effected.
For the less dense clusters it is actually the non-mass segregated clusters that lead to more encounter-induced loss, as here well-separated clusters lead to multiple gravitational foci, whereas in mass-segregated clusters basically only one gravitational focus exists.

Including stellar evolution in the simulations would lead to some additional mass loss. However, this additional mass loss would not set in for some time after gas expulsion. Depending on the actual massive star composition of the considered cluster, it will take a few Myr until the first star would explode as supernova. However, it is during these first few Myr when mass loss by gas expulsion and stellar interactions is most pronounced. For example, a cluster with \SFE\ =0.3 has completed at 5 Myr  $\sim$ 90\% of its total mass lost due to gas expulsion and $<$55\% of the mass lost due to stellar interactions. For the example of our model clusters LK1 (initial mass $\approx$ 18 000 \Msun),  the mass loss due to gas expulsion is $\sim$ 16 000 \Msun\  and that by encounters 500 \Msun . In comparison the mass loss due to the explosion of a single supernova is at most 150 \Msun\ . So the general conclusions of this paper will still hold if stellar evolution is included. If one includes stellar evolution, the bound cluster mass will continue to decline, also slowly, at times $>$ 20Myr. As a result the expansion will proceed but to a much lesser degree than during the first 20 Myr of the cluster development.

Neglecting primordial binaries could have more severe consequences. Unlike stellar evolution it influences the stellar dynamics right from the start. Although in above simulation binaries form very quickly by capture processes, their properties do not correspond to those of a primordial binary population. Simulations of ONC-like clusters, show that although capture processes lead to $\sim$ 20\% of binaries, however, they largely miss the very tight binary population with periastra smaller than 100 AU (Pfalzner \& Olczak 2007). Here, we find that in the clusters with $\rho$= 1 $\times$ 10$^3$ pc$^{-4}$ about $\sim $ 50\% of the ejected stars have been at some point part of a binary, whereas for denser clusters only   $\sim $ 30\% have been part of a binary.   

Energetic three-body interactions lead to ejections from the cluster and the connected mass loss. As close binaries are under-represented in above simulations,  ejections should be even more common in real clusters.  Performing a similar parameter study including binaries would be computationally very time consuming, especially for the densest systems. Therefore, we only performed the same kind of analysis for model LK1 with 30\% initial binaries. In this case the mass loss
due to encounters was 11\% compared to $\sim$ 6\% without primordial binaries. The situation might be different in the densest systems (model CK 1), where the higher densities might possibly favour rapid tight binary formation. This will require further investigation.

We showed that higher cluster concentrations (here Plummer vs. King W$_0$ = 9) lead to considerably higher mass remnant clusters if the SFE is in the range 0.2 $<$\SFE\ $<$ 0.4. Currently the  cluster profile at the on set of gas expulsion is observationally poorly restrained. As for such SFEs the outcome of gas expulsion seems so sensitive to the inner profile, it would highly desirable to obtain here better guidance from observations of massive clusters like, for example, Cyg OB2.

Here we assumed that the SFE throughout the cluster is constant. However, it has been argued that the SFE could be a function of the gas density or surface density, with higher gas densities leading to higher SFEs (Adams 2000, Parmentier \& Pfalzner 2013). For a cluster environment
this would result in a higher SFE in the cluster centre than at the outskirts. For the gas expulsion process this means a stronger binding of the central part and a higher bound fraction. In terms of the importance of stellar interactions, the higher central density compared to the constant SFE case leads to more interactions and mass loss due to stellar interactions should increase.

If one considers lower mass clusters, the assumption of instantaneous gas expulsion and
no sub-structuring are more critical. Lower mass clusters of the same size as their high-mass counterparts have lower stellar densities and one would expect that stellar interactions are much less important. However, in clusters of lower mass ($<$ 10$^3$ \Msun) gas expulsion probably takes much longer, because they contain fewer, if any, massive stars, driving the gas expulsion process. This results in slower cluster expansion leading to a slower decrease in cluster density and the cluster having more time to experience stellar interactions. Previous work showed that sub-structured initial conditions can lead to more massive remnant clusters. In lower mass clusters sub-structuring, thus locally enhanced densities leading to more stellar interactions, remains an issue for much longer times. Future investigations should address the relative importance of these two issues.

Here we start after gas expulsion has finished, however, the conditions in the cluster are probably very sensitive to what happened in the pre-gas expulsion phase.  In future a self-consistent treatment of these two phase would be the essential next step.

\section{Conclusion}

An extensive numerical parameter study of the bound cluster mass after gas expulsion as function of the star formation was performed, which largely confirm the qualitative results of previous work. However, what distinguishes the here presented results, is that it includes stellar interactions in a realistic way and shows the dependence of the bound fraction on the cluster density. This cluster density dependence is caused by additional mass loss due to ejections caused by stellar interactions between cluster members.

The reason why the effect of stellar interactions has been overlooked in the past, lies in a combination of the chosen cluster models and relatively large errors due to low sample sizes in the numerical simulations. Due to our large sample size this is the first quantitative description with an accuracy of $\sim$ 3\%. However, the most severe influence has the often chosen  single-mass representation of the stellar population instead of the full stellar mass spectrum present in a real cluster. In such models strong stellar interactions are generally under-represented and effects like gravitational focussing
(Pfalzner et al. 2006) completely absent.

Our results show that encounters alone can lead to mass loss of up to 20\% of the total cluster mass for the densest clusters typical in the Milky Way as, for example, the Arches cluster.   Even in less dense massive OB associations 5\% of the initial cluster mass is lost due to stellar interactions.  

The results of this numerical study are summarised in a simple fit formula (given in the appendix) for the bound mass fraction as a function of the SFE and the cluster density.  This density dependence is shallower than one would expect from 2-body relaxation processes because at cluster densities above 10$^4$ \Msun\ few-body interaction become important.

In this study no primordial binaries were included. This means that the here given high values of mass loss by ejection are actually lower limits, and could be even higher. We will address this problem in the near future.

\acknowledgement
We would like to thank the referee for the very constructive comments. Part of the simulations were performed at FZ J\"ulich under project number HKU14.

\appendix
\section{Fit formula for the bound fraction including density dependence}

In an accompanying paper (Pfalzner \& Kaczmarek, in preparation) we show how the effect of gas expulsion can be directly compared to observation for clusters more massive than 10$^4$ \Msun\ . There a fit formula for the total relative bound mass as a function of SFE and density is useful. Here presented results can be approximated by
\be \left( \frac{M_b}{M_{in}}  \right) = 
0.5 \left(1-0.7\arctan \left[ 12 (\epsilon_{SFE}-0.42) \right] \right) -  \left[\log(\rho)\right]^{3/2} \nonumber\\.\ee
 where the first term is the mass loss solely due to gas expulsion and the second term that due to encounter. If the cluster density is so low, that stellar encounters play no role, meaning for clusters with $\rho_0 < $ 10 \Msun pc$^{-3}$, the mass loss can be approximated by just using the first term. 

\begin{figure}[t]
\includegraphics[width=0.5\textwidth]{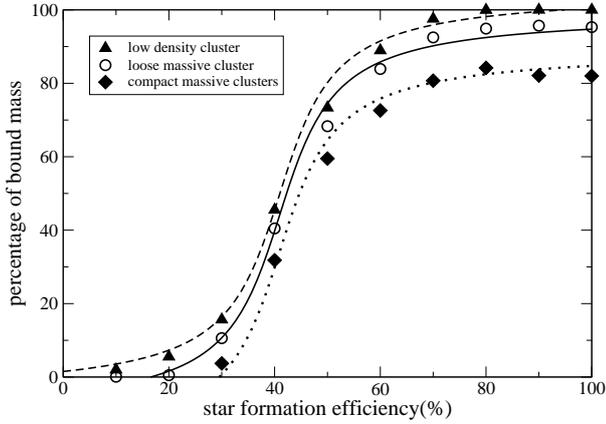}
\caption{The bound mass fraction after 20 Myr as a function of the SFE. The results from model LK1  are represented by circles and those of model CK1 by diamonds. Here the fits according to Eq. (A1) (represented as lines) and the numerical values (represented by the symbols) are shown for low-density clusters (dashed line, triangles), OB/leaky clusters (drawn line, circles) and starburst clusters (dotted line, diamonds).}
\label{fig:fit}
\end{figure}

A comparison between the numerical results for models LK1 (open circles) and CK1(filled diamonds) with the corresponding approximations is shown in Fig. \ref{fig:fit}.
Fig. \ref{fig:fit} includes a comparison between the numerical results for low density clusters (filled triangles) with this approximation (dashed line). The mean error of the fit formula compared to the numerical results is usually smaller than the numerical error, only for \SFE\ =0.5 the approximation gives a value that is 6.5\% smaller than the numerical result.

\end{document}